\documentclass{DISproc}

\begin{document}
\title{ Measurement of the forward energy flow in $pp$ collisions 
at $\sqrt{s}=7$~TeV with the LHCb detector}

\author{{\slshape Dmytro Volyanskyy\footnote{E-mail:
\href{mailto:Dmytro.Volyanskyy@mpi-hd.mpg.de}{Dmytro.Volyanskyy@mpi-hd.mpg.de}} and Michael Schmelling \\ 
on behalf of the LHCb collaboration}\\[1ex]
Max-Planck-Institut f\"ur Kernphysik, 69117 Heidelberg, Deutschland
 }

\contribID{xy}

\doi  

\maketitle

\begin{abstract}
 We present the results on the energy flow measured 
 with minimum-bias data collected by the LHCb experiment in $pp$ collisions 
 at $\sqrt{s}=7$~TeV for inclusive minimum-bias interactions, hard scattering processes
 and events with enhanced or suppressed diffractive contribution. 
 The measurements are performed in the pseudorapidity range $1.9<\eta<4.9$, 
 which corresponds to the main detector acceptance of the LHCb spectrometer.
 The data are compared to predictions given by the PYTHIA-based and 
 cosmic-ray Monte Carlo event generators, which model 
 the underlying event activity in different ways.
\end{abstract}

\section{Introduction}

The energy flow created in inelastic hadron-hadron interactions 
at large values of the pseudorapidity $\eta=-\ln\tan\theta/2$ 
with $\theta$\/ being the polar angle of particles w.r.t. the beam axis, 
is expected to be directly sensitive to the amount of parton radiation 
and multi-parton interactions (MPI)~\cite{PhysRevD.36.2019}. 
The latter mainly arise in the region of a very low 
$x =p_{\rm parton}/p_{\rm hadron} \rightarrow 0$, 
where parton densities are large so that the probability 
of more than a single parton-parton interaction 
per hadron-hadron collision is high. 
MPI represent a predominant contribution to
the soft component of a hadron-hadron collision, 
called the underlying event (UE). Its precise theoretical description 
still remains a challenge as MPI phenomenon is currently weakly known.

In this study, experimental results on the energy flow 
are compared to predictions given by the PYTHIA-based~\cite{Skands:2009zm,Clemencic:LHCbMC,P8}
and cosmic-ray Monte Carlo~(MC) event generators~\cite{Enterria:2011}, 
which model the UE activity in different ways. The analysis was performed using 
a sample of minimum-bias data collected by the LHCb experiment~\cite{LHCb} 
in $pp$ collisions at $\sqrt{s}=7$~TeV
during the low luminosity running period in 2010. 
The events were recorded using a trigger that has required 
the presence of at least one reconstructed track segment in the spectrometer. 
For a particular pseudorapidity bin with the width $\Delta\eta$, 
the total energy and total number of stable particles  $E_{tot}$ and $N_{part,\eta}$, 
the energy flow is defined as
\begin{equation}
\label{EF2}
 \frac{1}{N_{\rm int}} \frac{dE_{tot}}{d\eta} = \frac{1}{\Delta\eta}\left(\frac{1}{N_{\rm int}}\sum_{i=1}^{N_{part,\eta}}E_{i,\eta}\right)  \;,
\end{equation}
\noindent
where $N_{\rm int}$\/ is the number of inelastic $pp$\/ interactions and $E_{i,\eta}$ is the energy of an individual particle.


\begin{figure}[t!]
\centering
\includegraphics[width=0.475\textwidth]{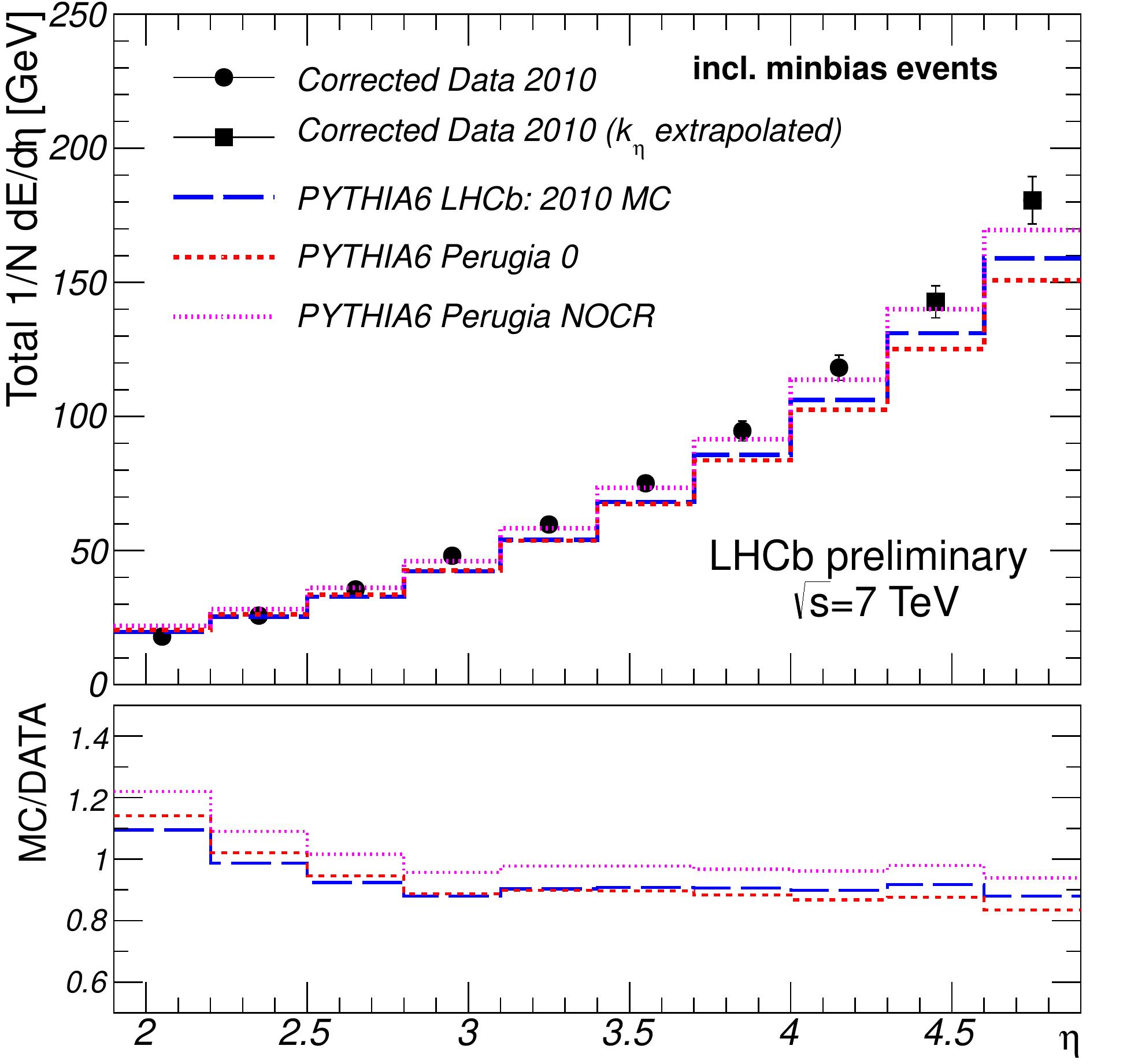}
\includegraphics[width=0.475\textwidth]{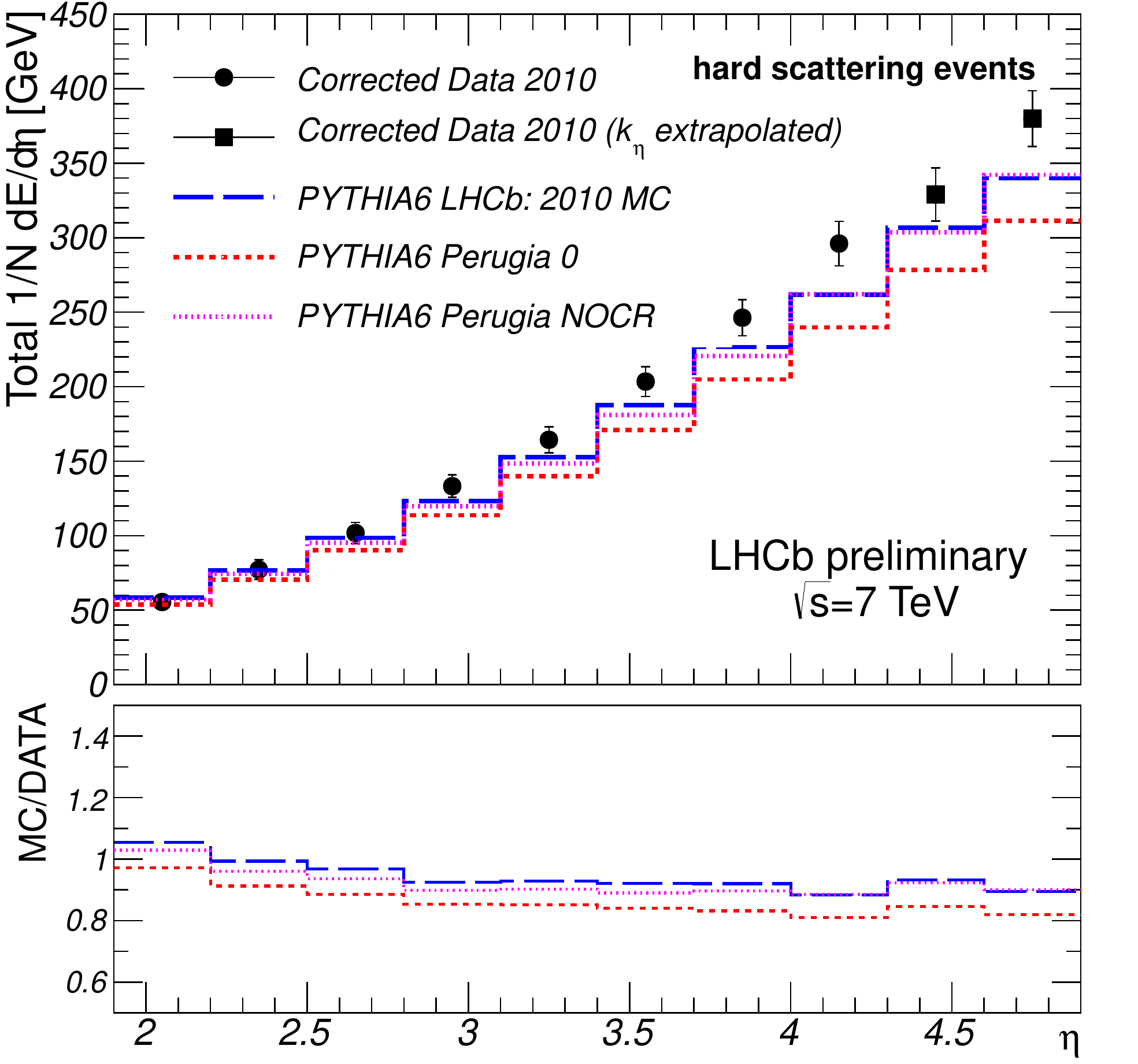}\\[5mm]
\includegraphics[width=0.475\textwidth]{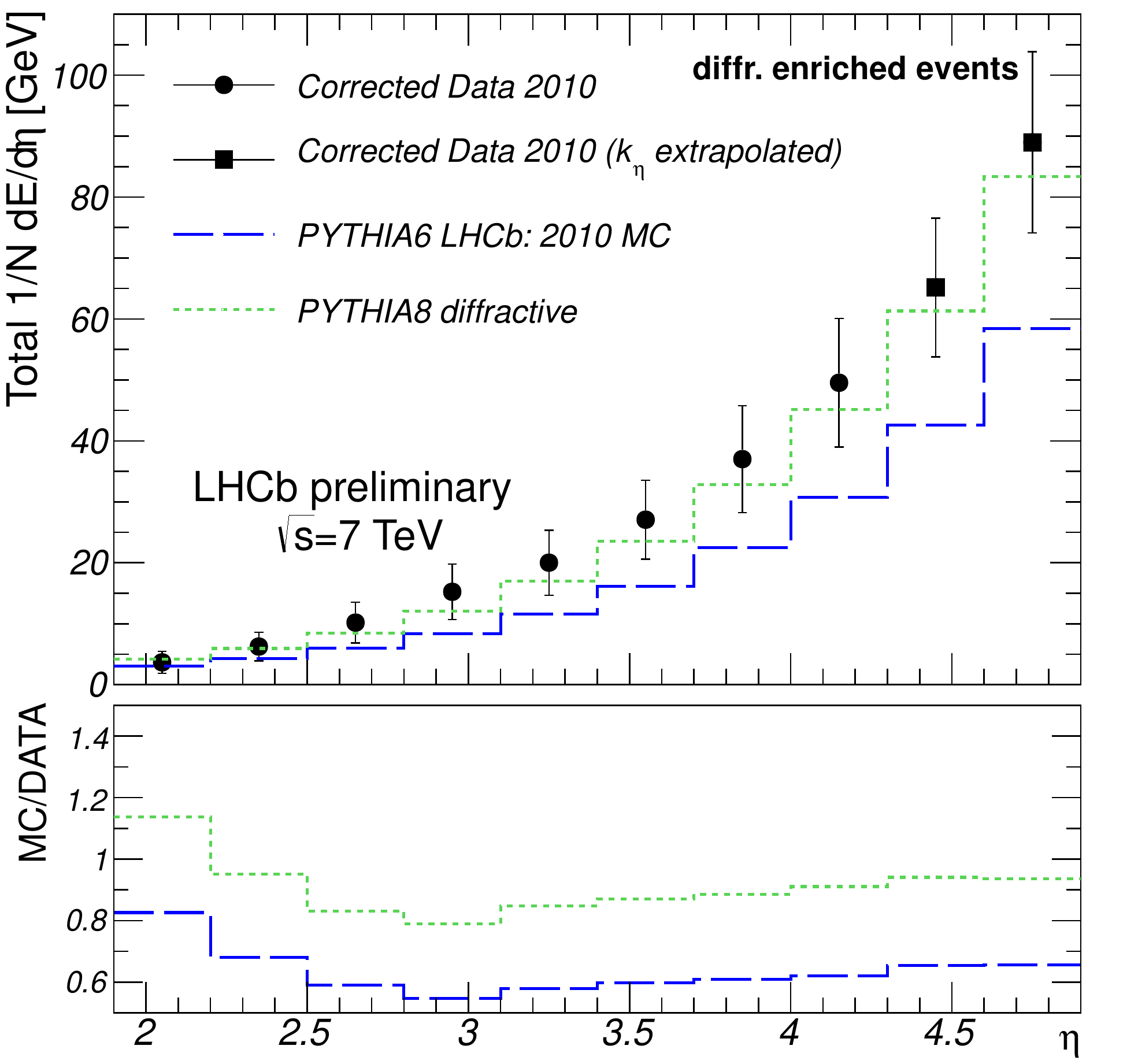}
\includegraphics[width=0.475\textwidth]{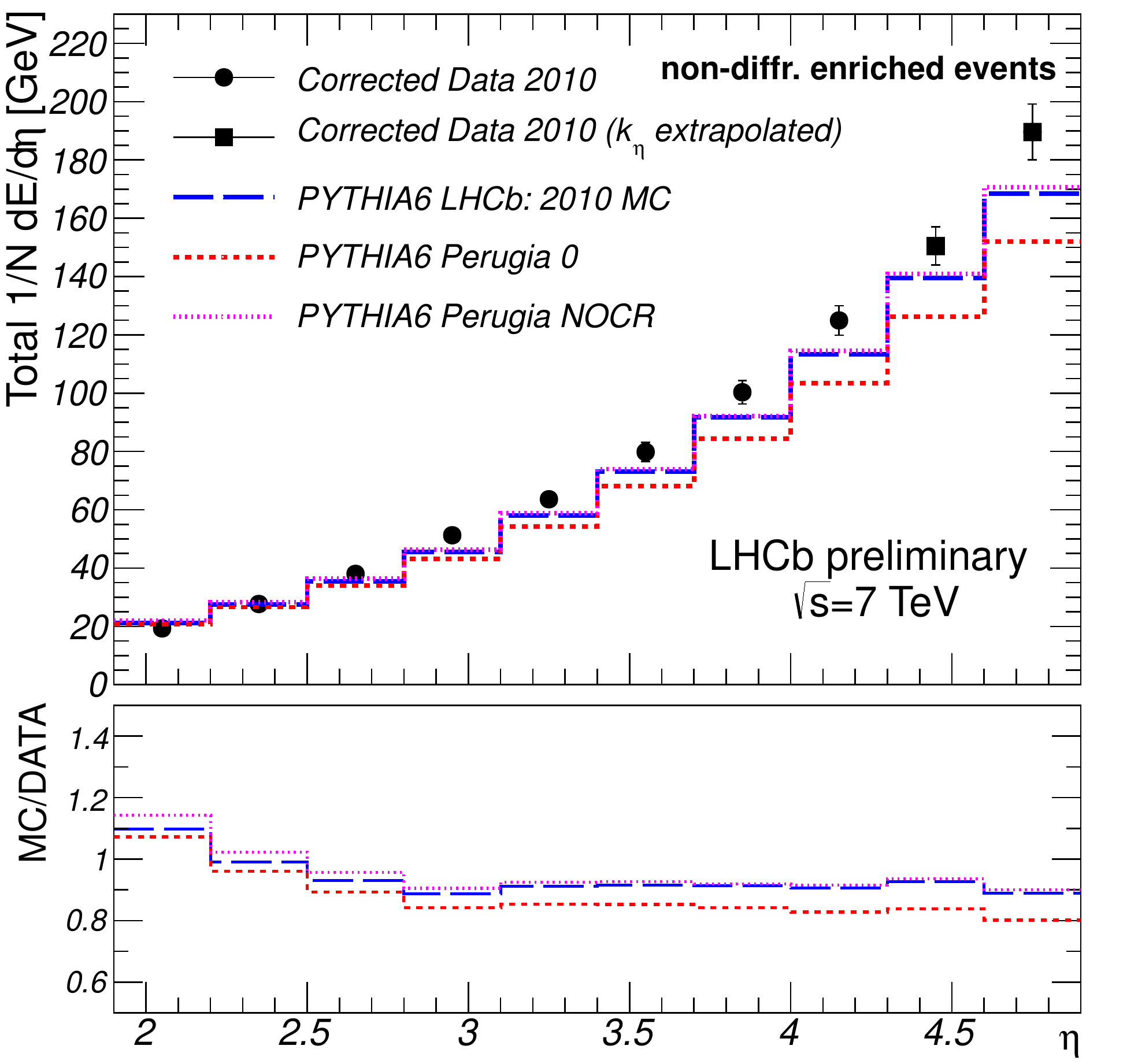}
\caption{\sl Total corrected energy flow obtained for all event 
             classes under consideration. The measurements are indicated 
             by points with error bars representing the systematic uncertainties, 
	     while the generator level predictions given by the PYTHIA-based models 
             are shown as histograms. 
             The ratios between the model predictions and corrected data are demonstrated in addition. 
        }
\label{fig:corrTotal} 
\end{figure}
\begin{figure}[h!]
\centering
\includegraphics[width=0.475\textwidth]{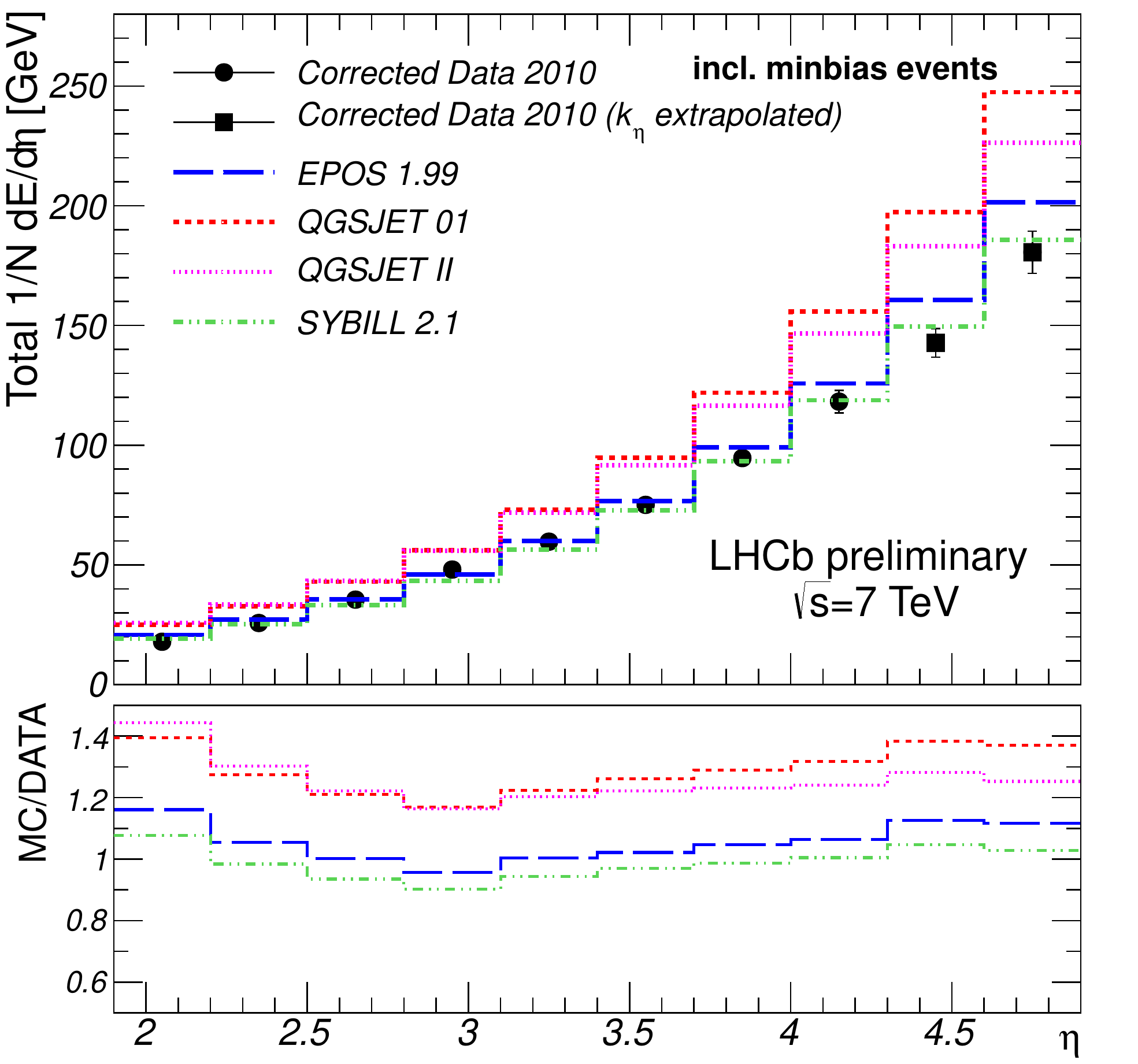}
\includegraphics[width=0.475\textwidth]{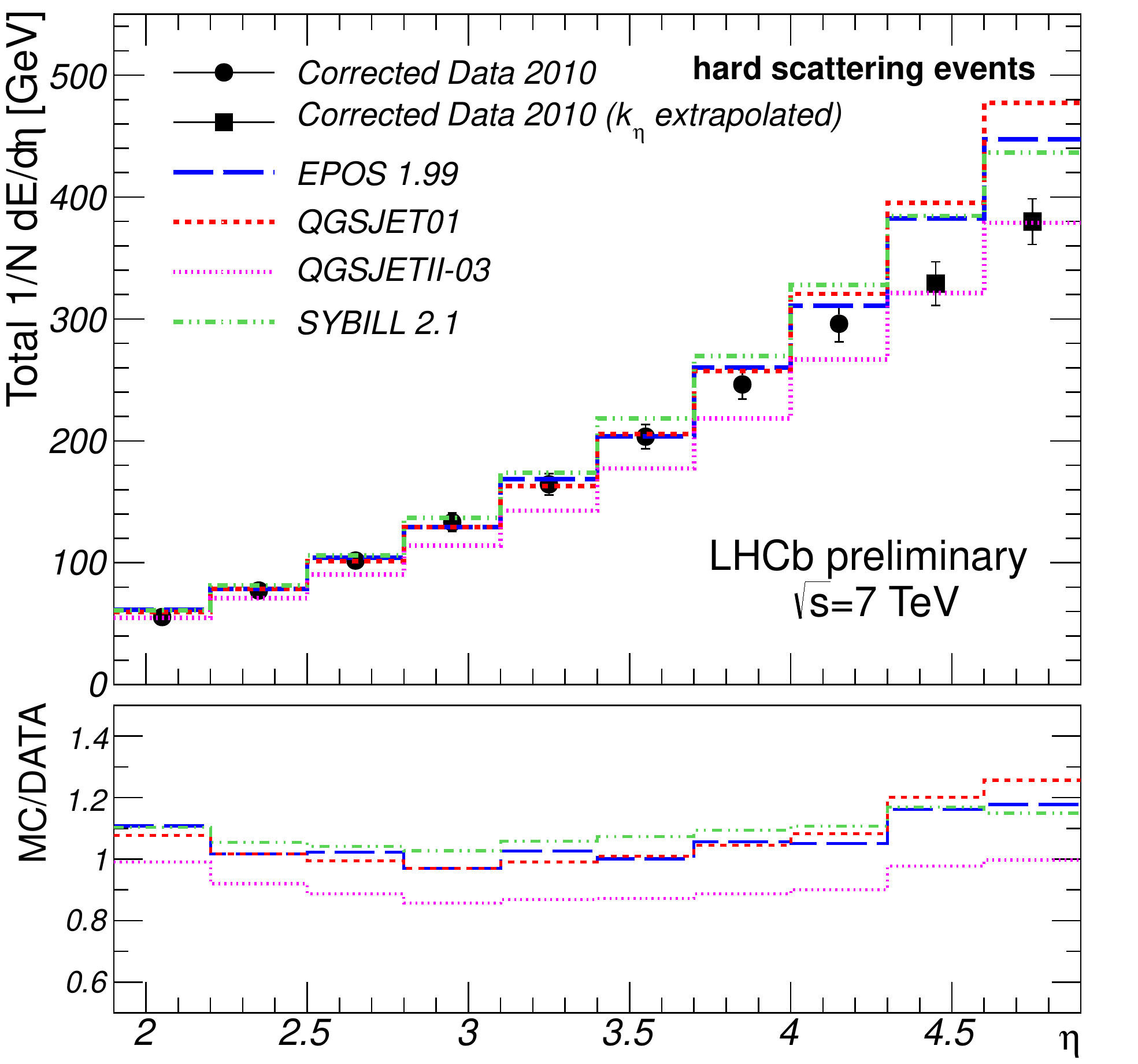}\\[5mm]
\includegraphics[width=0.475\textwidth]{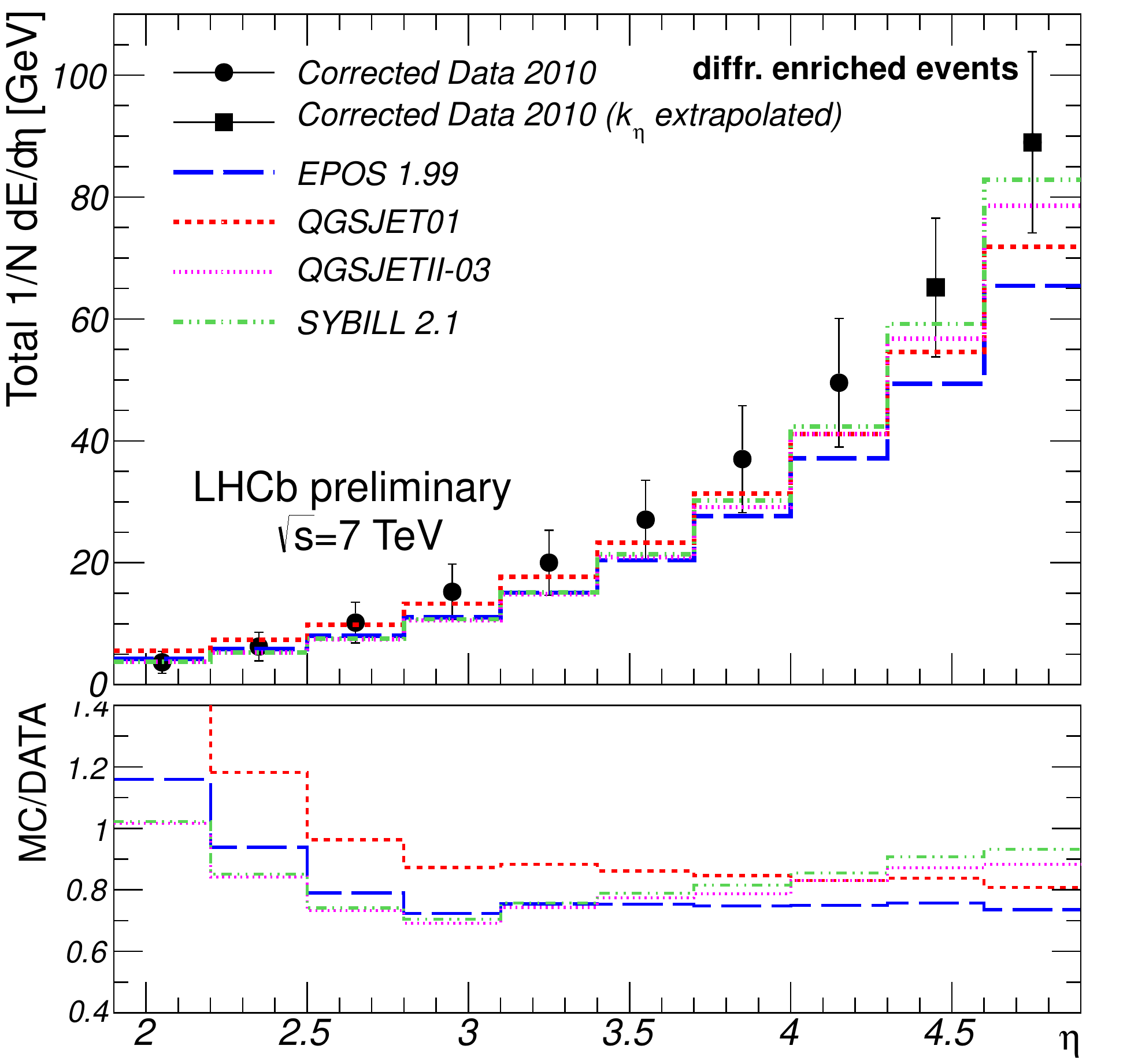}
\includegraphics[width=0.475\textwidth]{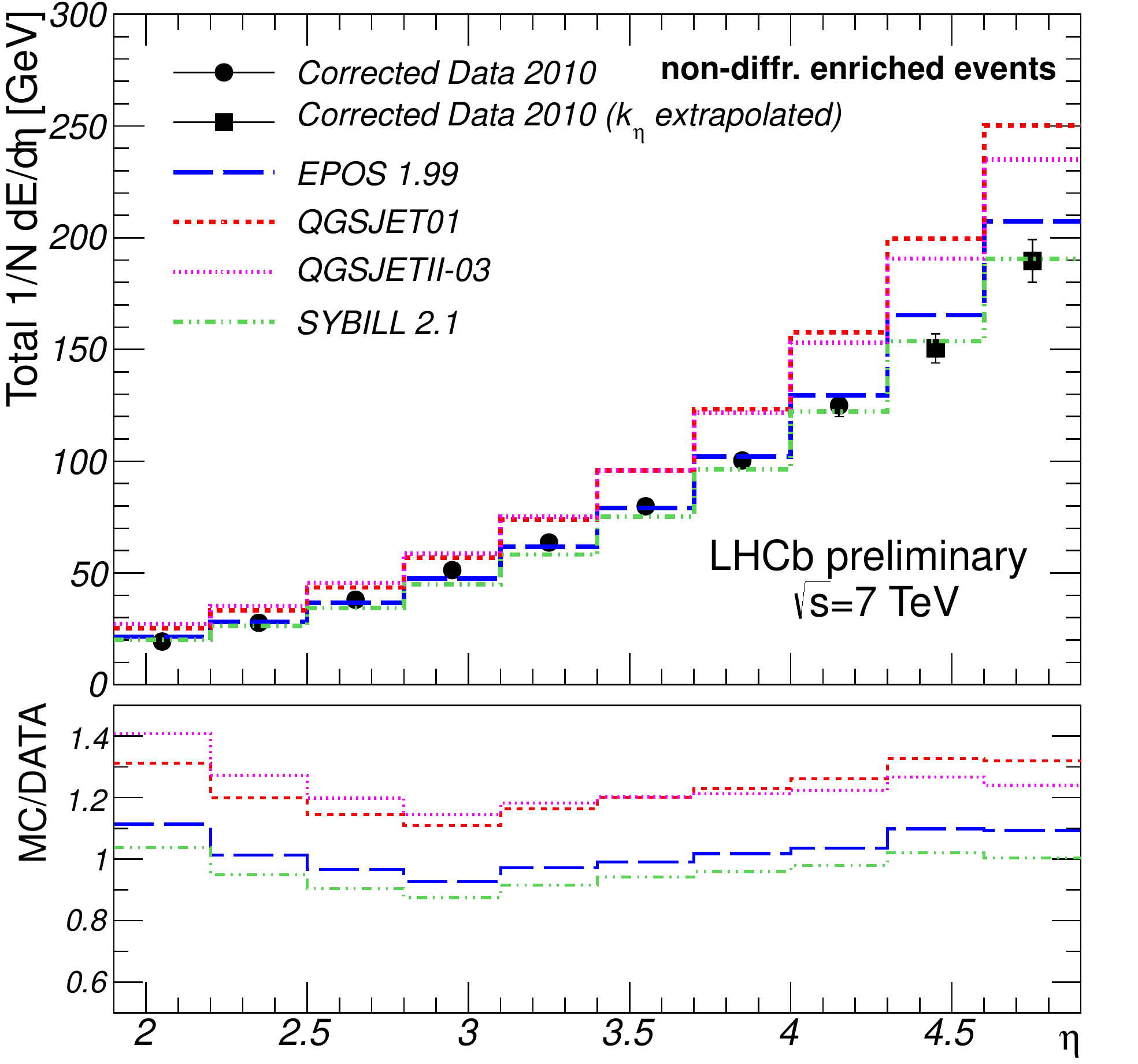}
\caption{\sl Total corrected energy flow obtained for all event 
             classes under consideration. The measurements are indicated by points 
             with error bars representing the systematic uncertainties, 
             while the predictions given by the cosmic-ray interaction models 
             are shown as histograms. 
             The ratios between the model predictions and corrected data are demonstrated in addition.
	}
\label{fig:corrTotalc} 
\end{figure}
\section{Analysis Strategy}

In this analysis, the energy flow carried by the charged stable particles 
was measured using good quality reconstructed tracks traversing 
the full LHCb tracking setup with a momentum in the range $2<p<1000$~GeV/c and $1.9<\eta<4.9$ .
Particle identification was not used in this study.
Instead, the energy was simply estimated from 
the reconstructed momentum. 
The reconstructed charged energy flow was corrected for detector effects 
using the average of correction factors obtained from various MC models
as the ratio of the predictions at generator and detector levels for each $\eta$\/ bin.
The total energy flow was estimated from the corrected charged one 
by using a data-constrained MC estimate of the neutral component. 
For the two highest $\eta$ bins 
the data-constrained measurements of the neutral energy flow 
were extrapolated from the more central region 
as the LHCb Electromagnetic Calorimeter has no detection coverage 
for that region of phase space.  

In order to probe various aspects of multi-particle 
production in high-energy hadron-hadron collisions, the measurements were performed 
for the following four event classes: inclusive minimum-bias (containing 
at least one track with $p>2$~GeV/c in $1.9<\eta<4.9$), 
hard scattering (having at least one track with $p_{\rm T}>3$~GeV/c in $1.9<\eta<4.9$), 
diffractive and non-diffractive enriched interactions. The last two event types 
were selected among the inclusive minimum-bias ones requiring 
the absence and presence of at least one backward track reconstructed by the LHCb Vertex Locator 
in $-3.5<\eta<-1.5$, respectively. 
A detailed description of the whole analysis procedure can be found in~\cite{LHCb-CONF-2012-012}.


\section{Results}
The computed total corrected energy flow is illustrated for every
event class in Fig.\,\ref{fig:corrTotal} and Fig.\,\ref{fig:corrTotalc} 
along with the PYTHIA-based and cosmic-ray model predictions, respectively. 
As can be seen, the energy flow increases with the momentum transfer 
in an underlying $pp$\/ inelastic interaction. 
The development of the energy flow as a function of $\eta$ 
is reasonably well reproduced by the MC models. 
Nevertheless, the PYTHIA-based generators underestimate
the corrected data at large $\eta$, while most of the
cosmic-ray interaction models overestimate it, except for 
the diffractive enriched events. The predictions given by the SIBYLL~2.1
cosmic-ray generator~\cite{SIBYLL} for the inclusive minimum-bias and
non-diffractive enriched events provide the best description of the
corresponding energy flows across the entire $\eta$ range of the measurements. 
In the forward region the total uncertainties for most
of the event classes are around $5\%$. For the diffractive enriched
events, the uncertainties are about $3$\/ times larger mainly because
of the strong model dependency of the correction factors. None
of the MC models used in this analysis are able to describe 
the energy flow measurements for all event classes that have been 
studied. It follows that the results obtained in this analysis can be used to improve 
the existing MC models by further constraining the parameters
describing the partonic stage of high-energy hadronic interactions,
diffractive particle production and hard QCD processes.

\section{Acknowledgements}

We are thankful to Colin Baus and Ralf Ulrich from the Karlsruhe Institute 
of Technology for providing the predictions of the cosmic-ray MC generators. 


{\raggedright
\begin{footnotesize}



\end{footnotesize}
}


\end{document}